\documentclass[journal = nalefd, manuscript = letter]{achemso}
\usepackage{graphicx,epsf,epsfig}
\usepackage{dcolumn}
\usepackage{bm}
\usepackage{color}
\usepackage{ulem}
\usepackage{subfigure}
\usepackage{upgreek}

\title{Self-aligned nanoscale SQUID on a tip}
\author{Amit Finkler}
\email{amit.finkler@weizmann.ac.il}
\author{Yehonathan Segev}
\author{Yuri Myasoedov}
\author{Michael L. Rappaport}
\author{Lior Ne'eman}
\author{Denis Vasyukov}
\author{Eli Zeldov}
\affiliation{Department of Condensed Matter Physics, Weizmann Institute of Science, Rehovot 76100, Israel}
\author{Martin E. Huber}
\affiliation{Departments of Physics and Electrical Engineering, University of Colorado, Denver, CO 80217, USA}
\author{Jens Martin}
\author{Amir Yacoby}
\affiliation{Department of Physics, Harvard University, Cambridge, MA 02138, USA}
\begin{document}

\begin{abstract}
A nanometer-size superconducting quantum interference device (nanoSQUID) is fabricated on the apex of a sharp quartz tip and integrated into a scanning SQUID microscope. A simple self-aligned fabrication method results in nanoSQUIDs with diameters down to 100 nm with no lithographic processing. An aluminum nanoSQUID with an effective area of 0.034 $\upmu \textrm{m}^2$ displays flux sensitivity of $1.8\times 10^{-6}\ \Phi_0/\sqrt{\textrm{Hz}}$ and operates in fields as high as 0.6 T. With projected spin sensitivity of 65 $\mu_B/\sqrt{\textrm{Hz}}$ and high bandwidth, the SQUID on a tip is a highly promising probe for nanoscale magnetic imaging and spectroscopy.
\end{abstract}
\maketitle

Imaging magnetic fields on a nanoscale is a major challenge in nanotechnology, physics, chemistry, and biology. One of the milestones in this endeavor will be the achievement of sensitivity sufficient for detection of the magnetic moment of a single electron \cite{Rugar2004}. There are three main technological challenges: fabrication of a sensor with a high magnetic flux sensitivity, reduction of the size of the sensor, and the ability to scan the sensor nanometers above the sample. Superconducting quantum interference devices (SQUIDs) have the highest magnetic field sensitivity, but their loop diameter is usually many microns. Much effort has been devoted recently to the development of nanoSQUIDs, which have shown very promising flux sensitivity \cite{Kirtley2009, huber:053704, Nb_nanoSQUID, hao:192507, Lam2006, Foley2009, CleuziouJ.-P.2006}. Most of these devices, however, are based on planar technology using lithographic or focused ion beam (FIB) patterning methods \cite{huber:053704, Granata2008, hao:192507, Nb_nanoSQUID, Foley2009, Lam2006, CHWu2008, CleuziouJ.-P.2006, Girit2009}; the large in-plane size of the devices precludes bringing the SQUID loop into sufficiently close proximity to the sample (due to alignment issues) to scan it with optimal sensitivity. Recently, a terraced SQUID susceptometer was developed that is based on a multilayered lithographic process combined with FIB etching. This device includes a 600 nm pickup loop which can be scanned 300 nm above the sample surface \cite{koshnick:243101}. Here we present a simple method for the  self-aligned fabrication of a DC nanoSQUID on a tip with effective diameter as small as 100 nm that can be scanned just a few nm above the sample.

We have fabricated several SQUID-on-tip (SOT) devices of various sizes. A quartz tube of 1 mm outside diameter is pulled to a sharp tip with apex diameter that can be controllably varied between 100 and 400 nm. The fabrication of the SOT consists of three ``self-aligned'' steps of thermal evaporation of Al, as shown schematically in Fig. 1a. In the first step, 25 nm of Al are deposited on the tip tilted at an angle of -100$^\circ$ with respect to the line to the source. Then the tip is rotated to an angle of 100$^\circ$, followed by a second deposition of 25 nm. As a result, two leads on opposite sides of the quartz tube are formed, as shown in Fig.\ 1b. In the last step 17 nm of Al are evaporated at an angle of 0$^\circ$, coating the apex ring of the tip. The two areas where the leads contact the ring form ``strong'' superconducting regions, whereas the two parts of the ring in the gap between the leads, indicated by arrows in Fig. 1c, constitute two weak links, thus forming the SQUID. The resulting nanoSQUID requires no lithographic processing, its size is controlled by a conventional pulling procedure of a quartz tube, and it is located at the apex of a sharp tip that is ideal for scanning probe microscopy.

The studies were carried out at 300 mK, well below the critical temperature $T_c \approx 1.6$ K of granular thin films of aluminum in our deposition system. Instead of the commonly used current bias, the SOT was operated in a voltage bias mode, as shown schematically in the inset to Fig.\ 2. We use a low bias resistance $R_b$ of about 2 $\Omega$ and the SOT current, $I_\mathrm{SOT}$, is measured using a SQUID series array amplifier (SSAA) working in a feedback mode \cite{Martinis, Huber-2001, hao:192507}. $R_s$ is a parasitic series resistance.

The resulting $I-V$ characteristics display several interesting features, as shown in Fig.\ 2. First, the advantage of our SOT and the voltage bias setup is that there is no hysteresis, which avoids the need for complicated pulsed measurements \cite{Hasselbach-2000}. Second, we observe a large negative differential resistance over a wide range of biases. This behavior is consistent with the Aslamazov-Larkin model of a single Josephson junction \cite{Aslamazov-1969} if the voltage bias circuit of Fig.\ 2 is taken into account.
Third, small SQUIDs often have a small modulation of the critical current with field \cite{Nb_nanoSQUID, Lam2006}. Our SOT, in contrast, shows very pronounced $I_c(H)$ modulation as seen in Fig.\ 2. Finally, the $I-V$ characteristics show fine structure at high biases, e.g., the 25 mT curve in Fig.\ 2, which apparently results from resonances, the exact nature of which requires further investigation.

Figure 3a shows $I_\mathrm{SOT}(V_\mathrm{in},H)$ plots displaying very pronounced quantum interference patterns with a period of 60.8 mT, corresponding to an effective SQUID diameter of 208 nm. The modulation of the critical current is large, $I_c^\mathrm{max}/I_c^\mathrm{min} = 1.67$, and in addition a large asymmetry between negative and positive biases is observed. Due to this asymmetry, the interference patterns at negative and positive bias are almost out of phase. From a theoretical fit \cite{Tesche1977} to $I_c(H)$, shown in Fig.\ 3a by the dashed curves, we extract the following parameters: the critical currents of the two junctions $(1-\alpha)I_0 =$ 0.8 $\upmu$A and $(1+\alpha)I_0 =$ 2.4 $\upmu$A, where $I_0 = 1.6$ $\upmu$A, the asymmetry parameter $\alpha = 0.5$, and $\beta=2LI_0/\Phi_0=0.85$, where $L$ is the loop inductance and $\Phi_0 = h/2e$ is the flux quantum. The asymmetric interference patterns therefore arise from the difference in the critical currents of the two junctions. This asymmetry is in fact very advantageous for scanning probe applications since high sensitivity can be attained essentially at any field by an appropriate choice of the SOT bias polarity and voltage.

The almost optimal $\beta=0.85$ of the SOT implies a large inductance $L=549$ pH. For comparison, the calculated geometrical inductance of our loop is $L_g = \mu_0R\left(\log\frac{8R}{r} - 2\right) = 0.26$ pH, which is more than three orders of magnitude smaller. Here $R = 104$ nm is the loop radius and $r = 15$ nm is the radius of the loop wire. Our device is therefore governed by the kinetic inductance \cite{Cardwell2002} of the loop, $L_k=2\pi\mu_0\lambda_L^2 R/a$, due to its small dimensions. Here $a = 510$ nm$^2$ is the estimated cross sectional area of the loop, resulting in penetration depth $\lambda_L=$ 0.58 $\upmu$m. This $\lambda_L$ is much larger than the bulk value for Al but is quite plausible for very thin granular Al films \cite{PhysRev.168.444, Gershenson1982}.

Usually SQUIDs are operational only at very low fields. Remarkably, the SOT can operate over a very wide range of fields without a significant reduction in sensitivity. Figures 3b and 3c show substantial quantum oscillations at fields as high as 0.5 T, which provides a unique advantage for investigation of various systems. This special property of SOT apparently arises from the fact that all the device dimensions are very small and the thin superconducting leads along the quartz tube are aligned parallel to the applied field.

Figure 4 shows the spectral density of the flux noise of the SOT at various applied fields. Above a few tens of Hz the low frequency 1/f -like noise changes into white noise on the level of $3\times 10^{-5}$ to $1.8\times 10^{-6}\ \Phi_0/\sqrt{\textrm{Hz}}$ over a wide range of fields, which translates into a field sensitivity of $1.1\times 10^{-7}$ T. Our flux sensitivity is comparable to that of state of the art SQUIDs \cite{Kleiner-2004}, yet the area of the SOT loop is only 0.034 $\upmu \textrm{m}^2$, which is the smallest reported to date \cite{Nb_nanoSQUID, Granata2008}. The small size of SOT is highly advantageous for spin detection since spin sensitivity in units of $\mu_B/\sqrt{\textrm{Hz}}$ is given by
$$S_n=\Phi_n\frac{R}{r_e}\left(1+\frac{h^2}{R^2}\right)^{3/2},$$
where $R$ is the radius of the loop, $h$ is the height of the loop above the spin dipole, $r_e = 2.82 \times 10^{-15}$ m, and $\Phi_n$ is the flux sensitivity in $\Phi_0/\sqrt{\textrm{Hz}}$ \cite{koshnick:243101,Ketchen1989}. For $h<R$ we obtain spin sensitivity of 65 $\mu_B/\sqrt{\textrm{Hz}}$ for spins located in the center of the loop with an on-axis magnetic moment. In the SOT geometry, however, enhanced sensitivity could be achieved by imaging the spins near the perimeter of the loop \cite{Tilbrook2009}. In this case the sensitivity is mainly determined by the width of the weak link of about 30 nm rather than the diameter of the loop of 208 nm, leading to an estimated sensitivity of about 33 $\mu_B/\sqrt{\mathrm{Hz}}$.
Such sensitivity should allow imaging of the spin state of a single molecule \cite{SKHLam2008}, for example $\textrm{Mn}_{12}$-acetate ($m = 20 \mu_B$), with integration of a few seconds. Moreover, our smallest operating SOT has an effective diameter of 130 nm. Assuming that it has the same flux noise as the larger SOT, the estimated spin sensitivity could be increased to below 20 $\mu_B/\sqrt{\textrm{Hz}}$.

We have integrated the SOT into a scanning probe microscope operating at 300 mK in which the tip is glued to one tine of a quartz tuning fork. The frequency shift or the reduction in amplitude of the resonance peak of the tuning fork are used as a feedback mechanism for tip proximity to the sample surface \cite{Atia-1997, karrai:1842}. This method provides the possibility of simultaneous imaging of sample topography and of the local magnetic field. The amplitude of the tip oscillation is typically less than 1 nm, and hence does not degrade the spatial resolution of the magnetic imaging. As a test sample we used a 200 nm thick film of Al patterned into a serpentine structure. Figure 5c shows a topographical scan across two adjacent strips of the seprpentine using the tip and the tuning fork. A transport current of 2 mA at 510 Hz was applied to the sample and the self-induced magnetic field was measured by the SOT at various heights above the surface. The results are shown in Fig. 5a. The data are in good agreement with the theoretically calculated field profiles shown in Fig. 5b. A field as low as 1 $\upmu$T is readily measurable, which allows detection of the seprpentine signal from a distance of 10 $\upmu$m above the surface.
  
In summary, we have developed a simple method for fabrication of sensitive nanoSQUIDs on the apices of sharp tips and have incorporated them into a scanning SQUID microscope. A nanoSQUID with effective area of 0.034 $\upmu \textrm{m}^2$ operated at fields as high as 0.6 T and displayed flux sensitivity of $1.8\times 10^{-6}\ \Phi_0/\sqrt{\textrm{Hz}}$, which translates into on-axis spin sensitivity of 65 $\mu_B/\sqrt{\textrm{Hz}}$. By optimizing the SOT parameters, a further reduction in the noise can be expected, which, combined with a smaller loop diameter, could lead to a significant improvement in spin sensitivity. Compared to other SQUID technologies, the ability to image magnetic fields just a few nm above the sample surface renders the SQUID on tip a highly promising tool for study of quantum magnetic phenomena on a nanoscale.

\begin{acknowledgement}
We thank D.E. Prober, M.R. Beasley, I.M. Babich, and G.P. Mikitik for fruitful discussions. This work was supported by the European Research Council (ERC) Advanced Grant, by the Israel Science Foundation (ISF) Bikura grant, and by the Minerva Foundation.
\end{acknowledgement}

\newpage

\section*{Figure captions}
\begin{description}

\item[Fig.\ 1]
(a) Schematic description of three self-aligned deposition steps for fabrication of SOT on a hollow quartz tube pulled to a sharp tip (not to scale). In the first two steps, aluminum is evaporated onto opposite sides of the tube forming two superconducting leads that are visible as bright regions separated by a bare quartz gap of darker color in the SEM image (b). In a third evaporation step, Al is evaporated onto the apex ring that forms the nanoSQUID loop shown in the SEM image (c). The two regions of the ring between the leads, marked by the arrows in (c), form weak links acting as two Josephson junctions in the SQUID loop. The schematic electrical circuit of the SQUID is shown in the inset of (c).
\item[Fig.\ 2]
$I-V$ characteristics of the SOT at 300 mK at different applied fields. The inset shows the schematic measurement circuit. The SOT is voltage biased using a small bias resistor $R_b$ and the current $I_\mathrm{SOT}$ is measured using a SQUID series array amplifier (SSAA) with a feedback loop.
\item[Fig.\ 3]
(a) Quantum interference patterns of the SOT current $I_\mathrm{SOT}(V_\mathrm{in},H)$ at 300 mK at positive and negative voltage bias. The patterns are asymmetric both in field and bias and are almost out of phase for the two bias polarities. The dashed line shows a theoretical fit taking into account the difference in critical currents of the two weak links. (b) Quantum interference patterns at high fields up to 0.5 T. (c) Current oscillations $I_\mathrm{SOT}(H)$ at a constant bias $V_\mathrm{in}=103.5$ mV over a wide field range.
\item[Fig.\ 4]
Spectral density of the flux noise of the SOT at 300 mK and different applied fields. The inset shows $I_\mathrm{SOT}(H)$ at a constant bias $V_\mathrm{in} = 100$ mV with the fields, indicated by colored circles, for which the noise spectra are presented. The lowest white noise level is $1.8\times 10^{-6}\ \Phi_0/\sqrt{\textrm{Hz}}$. The mismatch at $10^4$ Hz is an instrumental artifact.
\item[Fig.\ 5]
(a) Scanning SOT microscope measurement of a superconducting serpentine. Shown are the magnetic field profiles at various heights above the serpentine carrying a 2 mA current at 510 Hz. (b) Theoretically calculated field profiles at comparable indicated heights. (c) Topography across two strips of the serpentine based on feedback from a quartz tuning fork operating in constant height mode. (d) Schematic cross section of the serpentine showing the direction of the current in each strip.

\end{description}
\newpage
\begin{figure}
 \includegraphics[angle=270, width=\textwidth, keepaspectratio, clip]{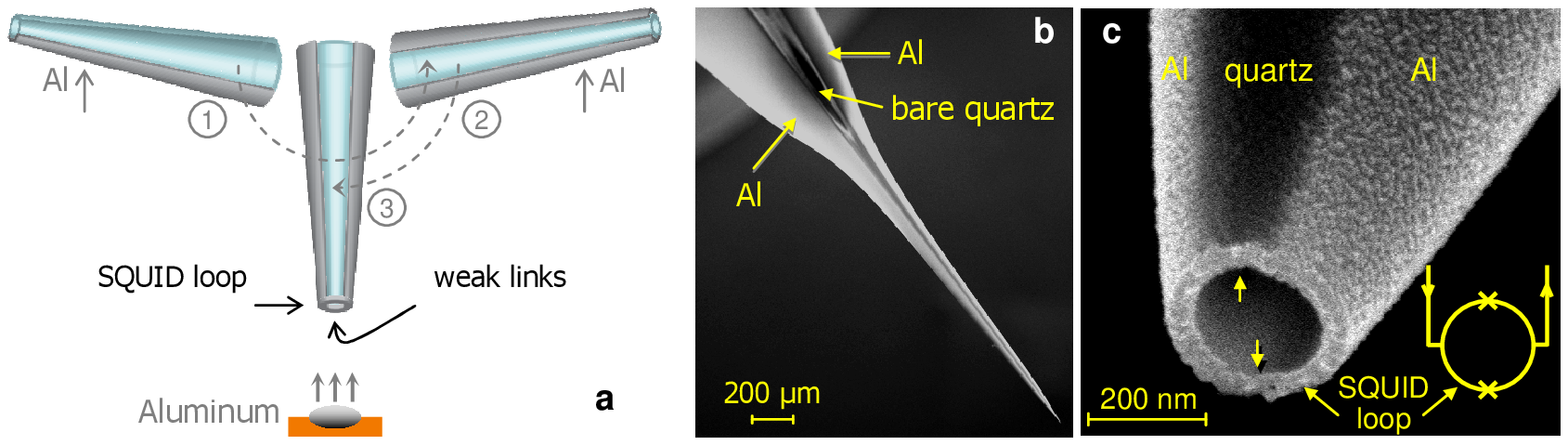}
\newline
\begin{center} 
  \LARGE Figure 1 
 \end{center}
\end{figure}

\newpage
\begin{figure}
 \includegraphics[width=\textwidth]{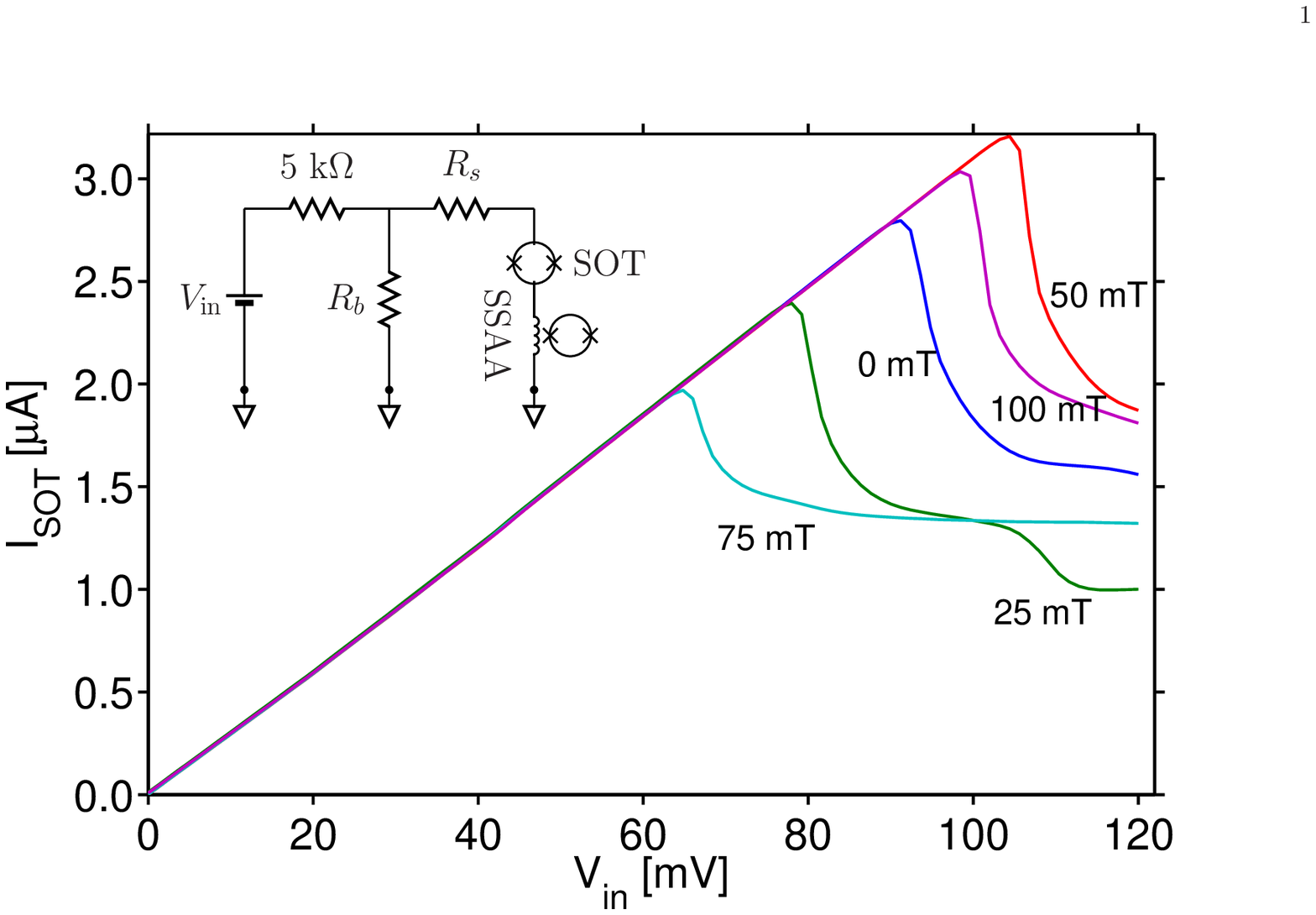}
 \begin{center} 
  \LARGE Figure 2 
 \end{center}
\end{figure}

\newpage

\begin{figure}
 \includegraphics[width = \textwidth]{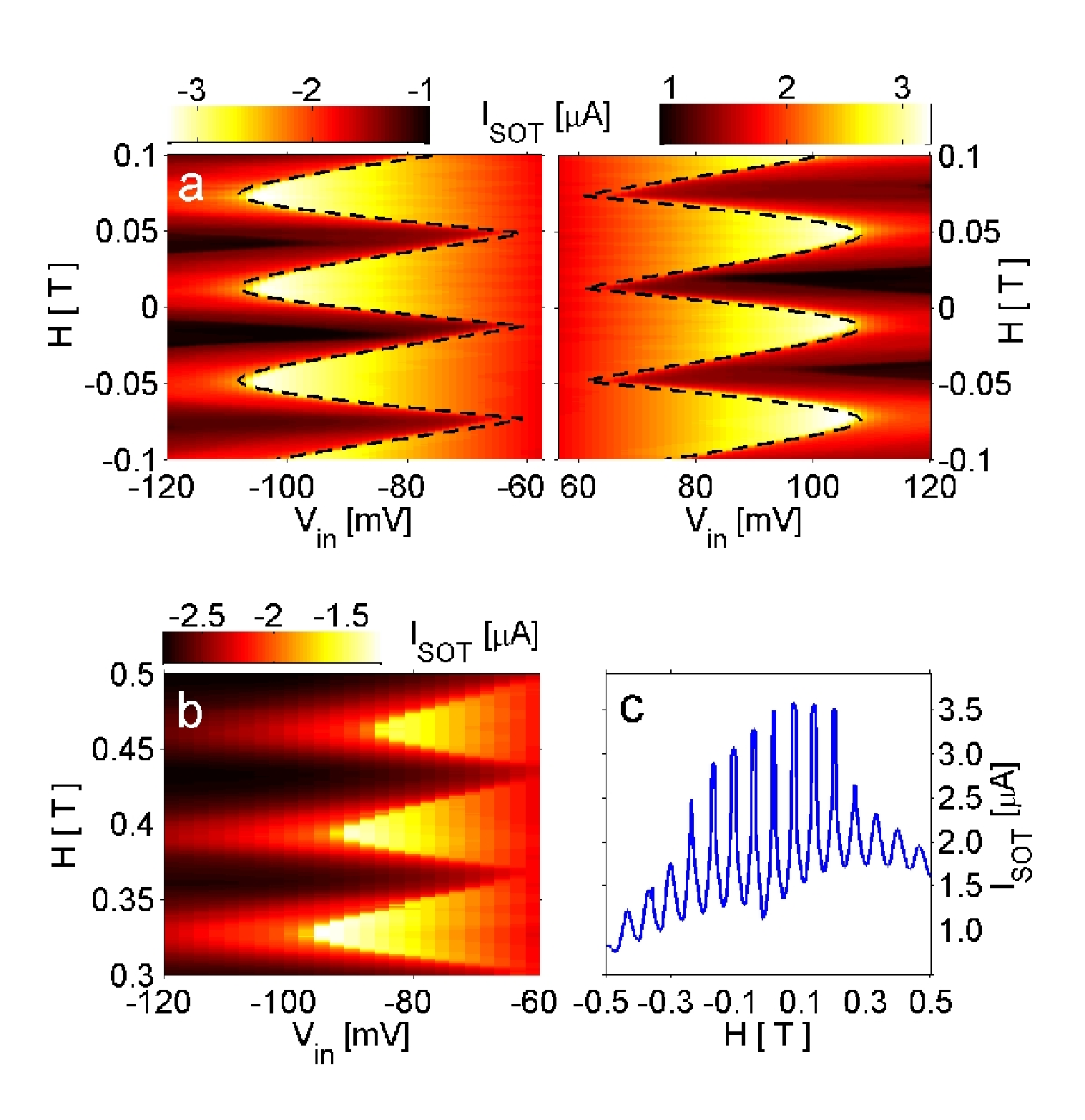}
 \begin{center} 
  \LARGE Figure 3 
 \end{center}
\end{figure}

\vspace{4in}
\newpage

\begin{figure}
 \includegraphics[width = \textwidth]{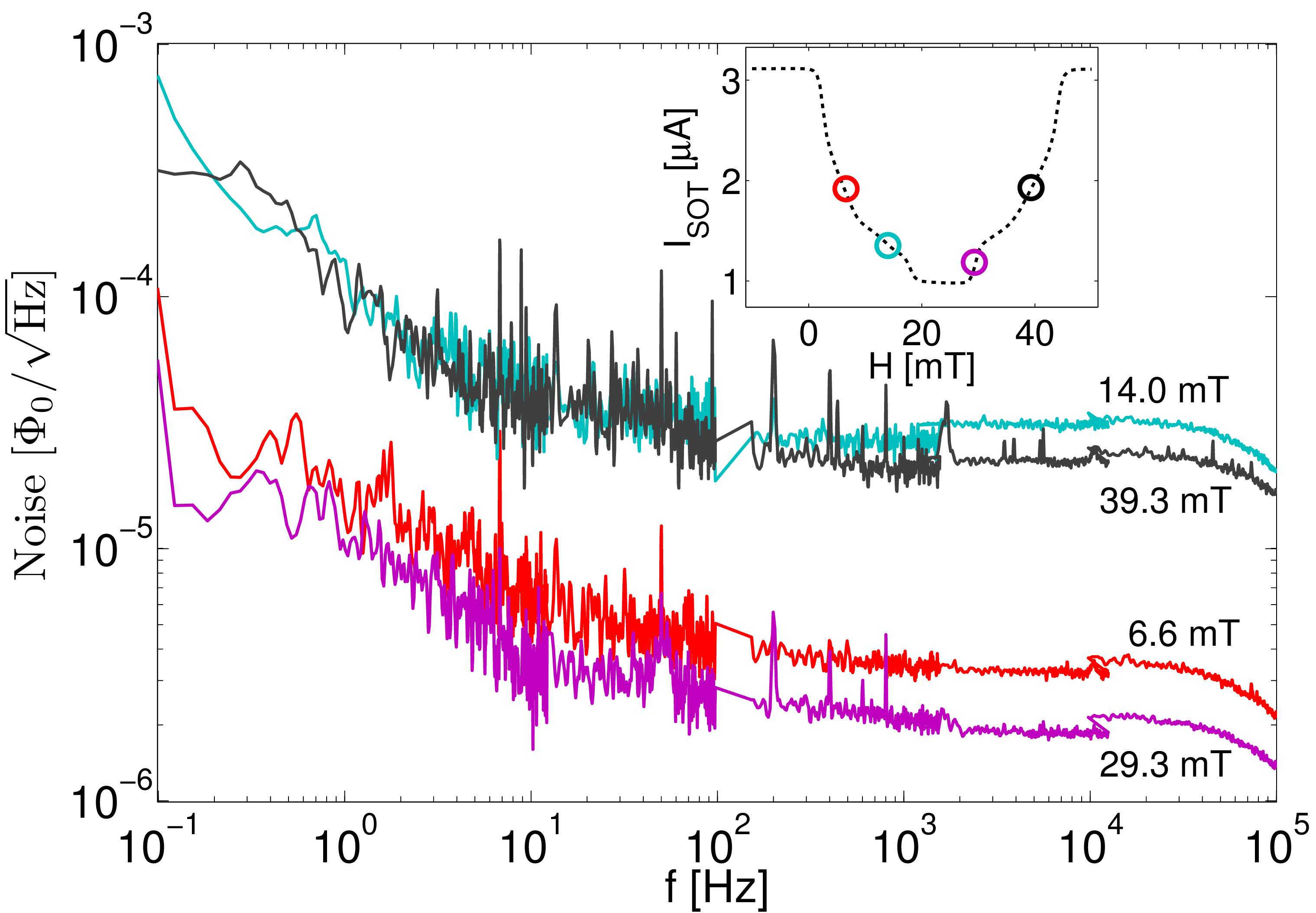}
 \begin{center} 
  \LARGE Figure 4 
 \end{center}
\end{figure}

\newpage

\begin{figure}
  \includegraphics[width = \textwidth]{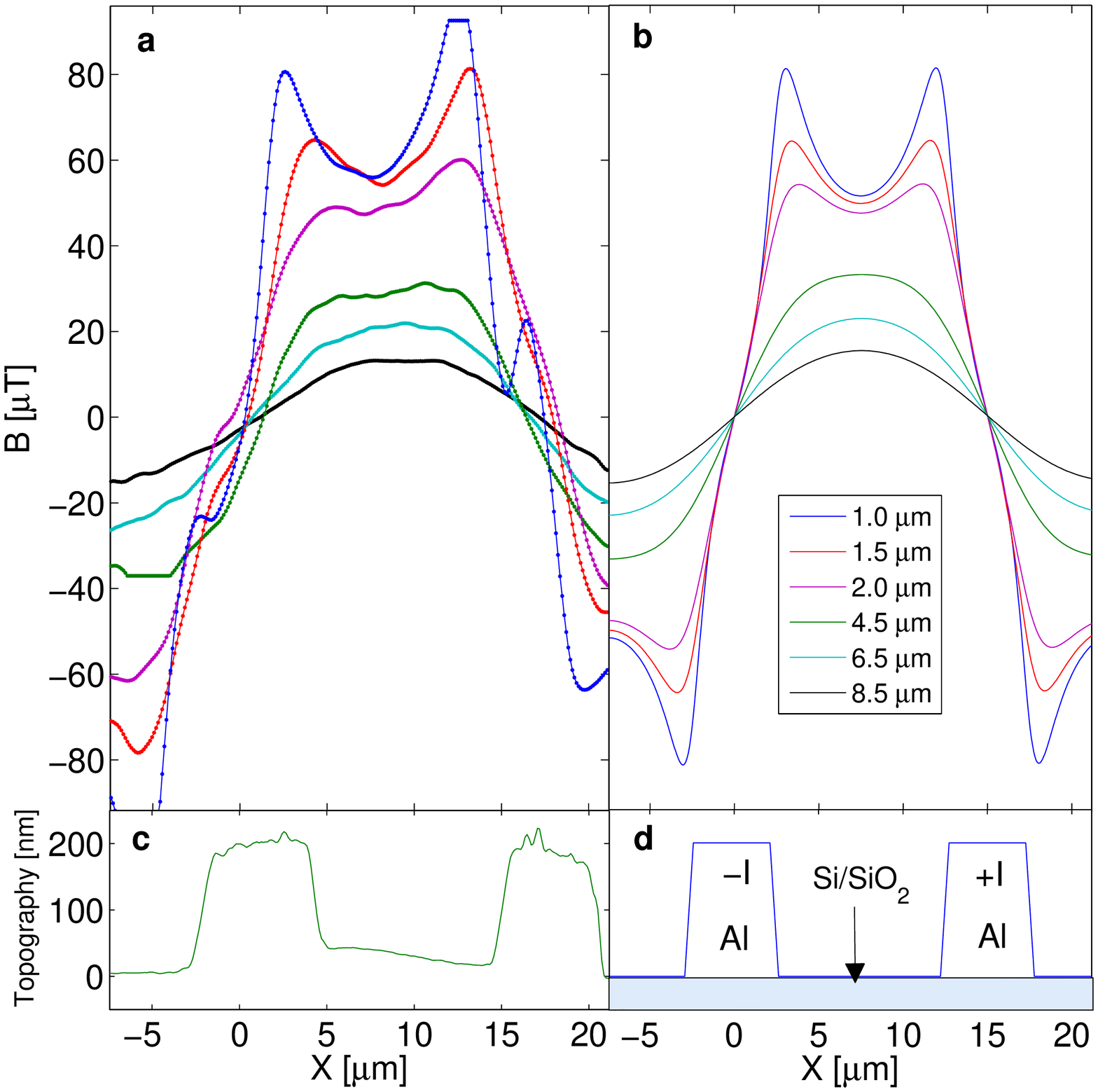}
 \begin{center} 
  \LARGE Figure 5 
 \end{center}
\end{figure}

\end{document}